# Existence of three-phase interlines on a cerium dioxide surface

Charles Osarinmwian

School of Chemical Engineering and Analytical Science, University of Manchester, Oxford Road, Manchester M13 9PL

**Abstract**

The three-phase interline described by a statistical continuum limit (i.e. quasi-boundary) has been postulated to gain a deeper insight into the reduction of $CeO_2$ to $CeO_{1.940}$ in a LiCl-KCl eutectic melt. Fabrication of a $CeO_2$ superstructure by a condensed-phase method provided a $CeO_2$ (111) surface at the nanoscale, which allowed the three-phase interline to be identified given previously reported quantum confinement effects in quasi-stoichiometric $CeO_2$ nanoparticles. Also, the $CeO_2$ superstructure displays the same crystal lattice planes as a bulk $CeO_2$ grain but the triply degenerate Raman-active peak of the grain is higher by a factor of ~ 2.5 with a wider full width at half maximum.

## 1. Introduction

An understanding of the reduction of $CeO_2$ to $CeO_{1.940}$ in air plasma and the oxidation of $CeO_{1.940}$ back to $CeO_2$ by heating in air at 500 °C for several hours [1] may require knowledge of three-phase interlines on the $CeO_2$ surface. Materials based on $CeO_2$ are used in the production and purification of hydrogen, various catalytic applications (e.g. automotive catalytic converters), and as a surrogate material for studying the properties $PuO_2$ in nuclear fuel technology. The characterization of oxygen vacancies on $CeO_2$ has led to greater understanding of the fundamental features of reactivity as well as the design of efficient supported catalysts [2]. These oxygen vacancies are known to localize excess electrons on a single cerium ion on a $CeO_2$ surface [3] which can be interpreted by a three-phase interpoint. This interpretation could enhance understanding of the mechanism for the reduction of $CeO_2$ to $CeO_{1.940}$. In nuclear fuel research, a corrosion test on fine $CeO_2$ powder in an alkali melt at 1050 °C for 40 h led to the statement that $CeO_2$ was inert in the melt [4]. However, a crystal structure analysis is required to give the implications relating to the reactive and thermal properties of the next generation of nuclear fuel [5, 6]. In this paper a condensed-phase method for fabricating $CeO_2$ at the nanoscale from a LiCl-KCl eutectic melt is reported, and the identification of three-phase interlines on a $CeO_2$ surface proposed.

## 2. Experimental

Nanoscale $CeO_2$ was synthesized by a condensed-phase method that involved mixing $CeO_2$ powder with LiCl and KCl powders. Powder was prepared in a eutectic composition by mixing 8.84 g pure LiCl (Alfa Aesar, > 98 %) with 11.16 g KCl (Aldrich, 99.0 %) [7]. The mixture was heated in air at 350 °C for 4 h and 500 °C for 2 h in an alumina crucible, and then rapidly quenched in air before grinding into powder. The $CeO_2$ powder (Alfa Aesar, 96 %) was mixed with the LiCl-KCl eutectic powder and then heated in air at 1050 °C for 40 h in an alumina crucible. At 40 h the sample was quenched in air, immersed in de-ionized water and then filtered to recover the solid residue. The dried residue was characterized by scanning electron microscope (JEOLJSM-6490LV) equipped with an energy-dispersive X-ray spectrometer (INCAx-act), XRD (Rigaku Miniflex), and a Bruker FRA 106/5 Raman Spectrometer with a coherent 500 mW laser.

## 3. Results and Discussion

Removal of moisture from pendular liquid bridges between $CeO_2$ nanoparticles (Fig. 1) forms a superstructure consisting mainly of solid bridges after the LiCl-KCl eutectic melt (melting point: 352 °C) is fully vaporized. In general, the bond strength due to the presence of a pendular bridge arises from the pressure drop over a particle surface produced by the curvature of the liquid meniscus as well as the interfacial tension exerted by the liquid along the wetted perimeter [8]. Related work, using a NaCl-KCl eutectic melt, fabricated metal-nanoparticle superstructures by the self-assembly of liquid metal and nanoparticles by pendular bridges [9]. The underlying mechanism of this self-assembly was unraveled by analysis of the Gibbs free energy. Figure 1 shows that the surfaces of $CeO_2$ nanoparticles appear atomically sharp, and clean without visible contaminants. Also, there are probably void spaces between these nanoparticles with very strong solid bridges between them that form during sintering at 1050 °C followed by rapid quenching in air.

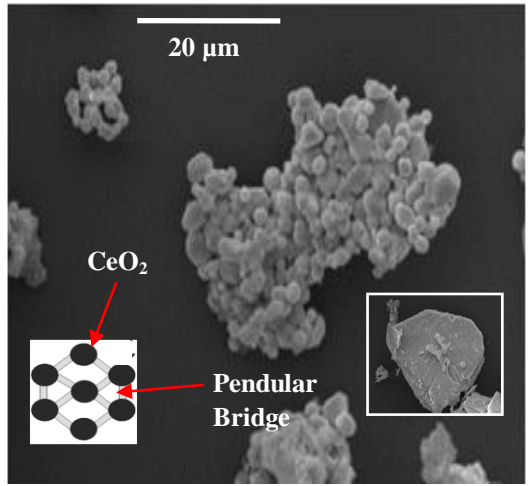

**Figure 1: Microstructure of a $CeO_2$ superstructure that may contain pendular liquid bridges. These types of bridges are weaker than solid bridges. Inset: microstructure of a $CeO_2$ grain from the supplier (i.e. bulk $CeO_2$).**

The bulk $CeO_2$ grains are still present in the $CeO_2$ superstructure after processing. The Raman line positioned at 465.7 cm$^{-1}$ (Fig. 2a) corresponds to the triply degenerate $F_{2g}$ mode for a cubic fluorite crystal consisting of three fcc sublattices, one of $Ce^{4+}$ ions and two of nonequivalent $O^{2-}$ ions, where the $O^{2-}$ ions vibrate symmetrically around the central $Ce^{4+}$ ion [10]. Bragg peaks with Miller indices (111), (200), (220), (311) and (222) in room temperature XRD spectra support observations of the $CeO_2$ lattice (Fig. 2b). On the thermodynamically most stable (111) surface linear clusters of oxygen vacancies may form and strongly bind adsorbates [2] which contribute to pendular bridges during agglomeration in the melt (Fig. 1). Also, the incomplete reduction of $CeO_2$ may generate a certain amount of the fully reduced $Ce_2O_3$ [3]. This could lead to an insulating $CeO_2$ (formally $Ce^{4+}$ with an f$^0$ configuration: empty 4f band)/metallic $Ce_2O_3$ (formally $Ce^{3+}$ with an f$^1$ configuration: partially occupied 4f band)/melt three-phase interpoint (Fig. 2b) and therefore an insight into the mechanism of the reduction of $CeO_2$ to $CeO_{1.940}$. However, more detailed work may be required to detect the change in stoichiometry between $CeO_2$ and $CeO_{1.940}$ as they share the same fluorite structure without significant alteration in lattice parameter [1].

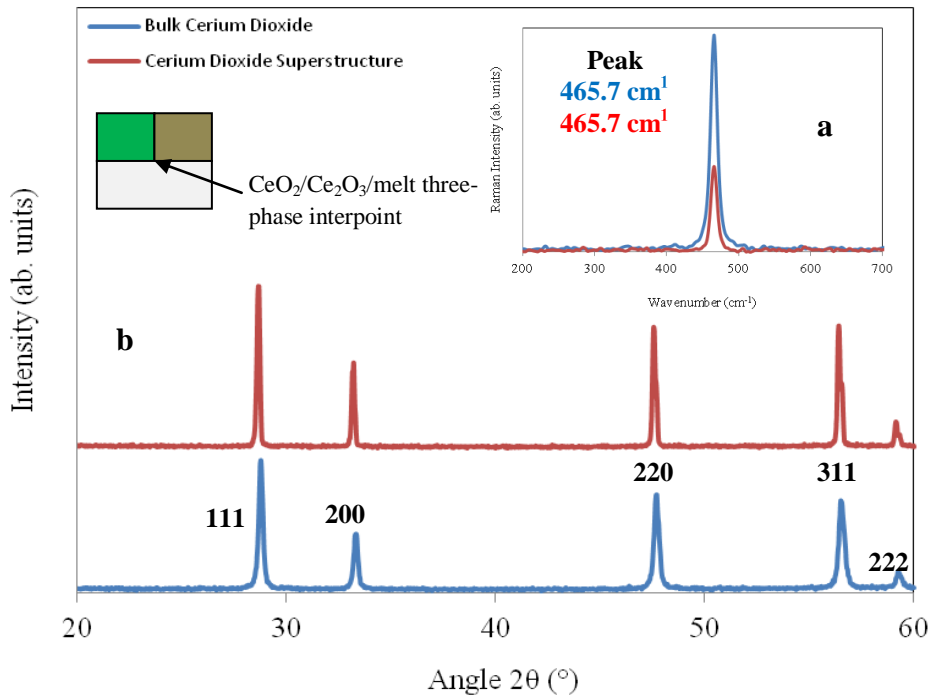

**Figure 2: Crystal structure of CeO$_2$.** (a) X-ray diffraction spectra for bulk CeO$_2$ and CeO$_2$ superstructure with Miller indices (hkl) identifying each surface. The existence of a CeO$_2$/Ce$_2$O$_3$/melt three-phase interpoint on a CeO$_2$ surface is shown. Colors: Green is CeO$_2$; Brown is Ce$_2$O$_3$; White is melt. (b) Symmetric Raman lines where the full width at half maximum for bulk CeO$_2$ is greater than the CeO$_2$ superstructure.

The formation of linear clusters of oxygen vacancies on the CeO$_2$ surface suggests that the CeO$_2$/Ce$_2$O$_3$/melt three-phase interline is linear (Fig. 3). Given that quantum confinement effects in the optical bandgap of quasi-stoichiometric cerium dioxide nanoparticles can be justified on the basis of a Ce$^{3+}$ related effect [11] a hypothetical length dimension of a CeO$_2$/Ce$_2$O$_3$/melt three-phase interpoint within the three-phase interline is assumed to be of the same magnitude as the wavelength of the electron wave function. This assumption reconciles the absence of a characteristic length scale of the statistical continuum limit [12] described for the three-phase interline by providing a classical length scale that neglects interpoint spacing. It is important to note that independent field variables are still conceived by a continuum boundary (applying the Bohr correspondence principle) and so field averages consist of integrals over the value of the field at each three-phase interpoint; not infinitely many integration variables in a classical continuum limit. The quantum shuffle diagram demonstrates that entropy in a classical continuum limit can only be known empirically but is a function of energy distribution in three-phase interpoints in a statistical continuum limit. Therefore, the classical continuum limit in the quantum shuffle diagram is equivalent to a single microscopic state available at zero Kelvin and so at zero entropy (i.e. third law of thermodynamics). This comparison leads to a nonsensical condition and so further supports a three-phase interline described in a statistical continuum limit on the $CeO_2$ surface.

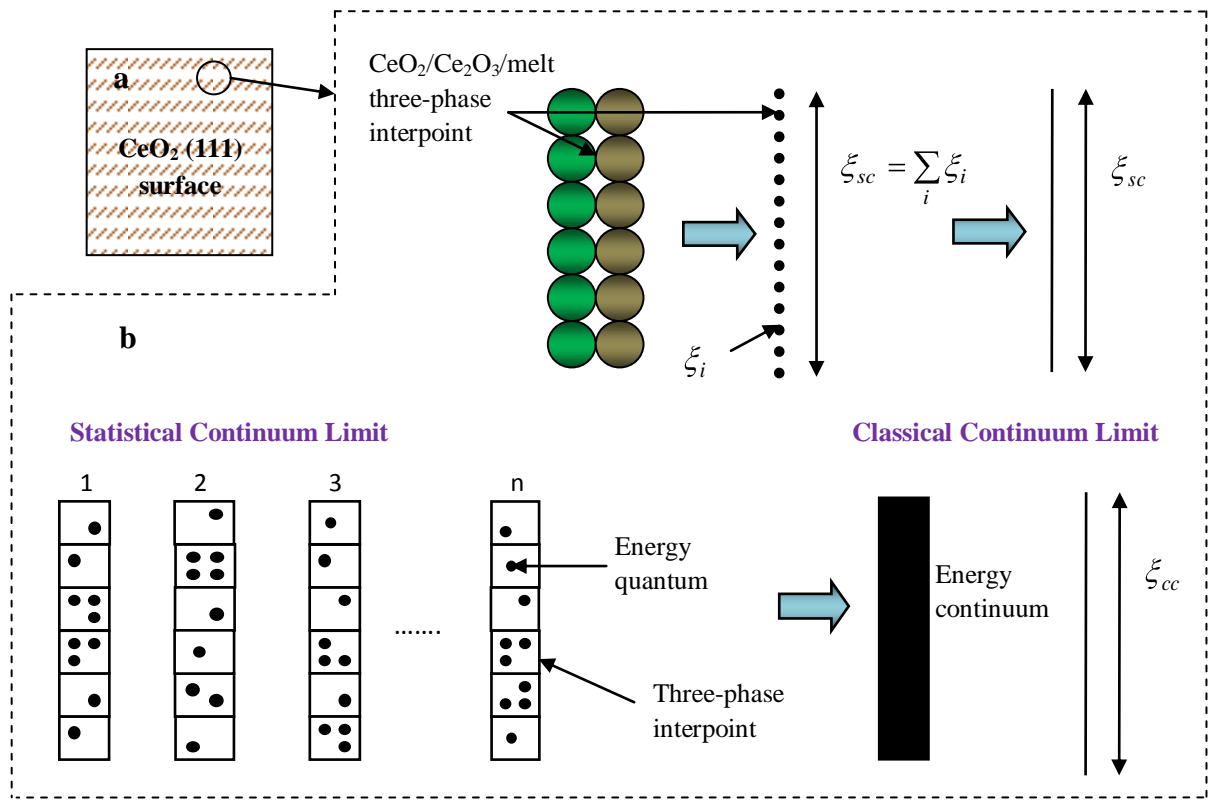

**Figure 3: (a) Confinement of $CeO_2/Ce_2O_3$/melt three-phase interlines on a $CeO_2$ (111) surface. (b) Presentation of the $CeO_2/Ce_2O_3$/melt three-phase interline in a statistical continuum limit with a hypothetical length dimension, $\xi$. Quantum shuffle diagrams indicate many ways of distributing energy in the three-phase interline ($W$) with reference to the Boltzmann equation. The entropy is zero when the number of ways of distributing energy in the three-phase interline is unity ($S = k_B \ln(1) = 0$) in a classical continuum limit.**

The three phase interline extends the new class of reaction occurring at a pinpoint [13] to multiple pinpoints (i.e. three phase interpoints) that are described by a statistical continuum limit, and provides a new perspective for describing the reduction at the surface of $CeO_2$ as well as the oxidation of carbide surfaces [14]. More importantly this concept represents a move from the paradigm of two-phase interface boundaries towards three-phase interlines as it can address the scientific challenge of relating singular macroscopic properties to interactions among very large numbers of microscopic states. In comparison, the triple phase boundary concept (extensively used in fuel cell literature) holds that the hydrogen oxidation reaction and the oxygen reduction reaction can only occur at confined spatial sites where electrolyte, gas, and electrically connected catalyst regions contact [15]. However, previous efforts to clearly delineate the nature and properties of the triple phase boundary into a continuum field theory has been challenging since the constitutive behavior of electron-electron interactions preclude reasonable descriptions at the process scale. However, the three-phase interline is able to convolute fundamental material properties and microstructure geometry into a detailed study of the electrochemical behavior of fuel cells and batteries, and provides a tool for the modeling and scale-up of production processes [16].

## 4. Conclusions

The existence of $CeO_2/Ce_2O_3$/melt three phase interlines have been postulated for the first time in order to gain a full understanding of the reduction of $CeO_2$ to $CeO_{1.940}$ in an ionic melt. The application of the three-phase interline in the statistical continuum limit has not only shown the ability to describe the quantum confinement effects at $CeO_2/Ce_2O_3$/melt three phase interlines but could replace the classical continuum limit used in a wide range of interfacial boundary problems in science and engineering. Further, the formation of a $CeO_2$ superstructure by the condensed-phase method used here could provide a simple method for controlling the production self-assembled nanoparticle superstructures.